\documentclass[conference]{IEEEtran}
\IEEEoverridecommandlockouts

\usepackage[nobreak]{cite}

\usepackage{setspace}

\usepackage{amsmath,amssymb,amsfonts}
\usepackage{algorithmic}
\usepackage{graphicx}
\usepackage{textcomp}
\usepackage{xcolor}
\def\BibTeX{{\rm B\kern-.05em{\sc i\kern-.025em b}\kern-.08em
    T\kern-.1667em\lower.7ex\hbox{E}\kern-.125emX}}
    
\usepackage[left=0.55in,right=0.55in,top=0.6in,bottom=0.6in]{geometry}

\newcommand{\defn}[1]{\textit{#1}}
\newcommand{\x}[0]{$\times$}

\usepackage[colorinlistoftodos,prependcaption,textsize=scriptsize]{todonotes}
\begin{document}

\title{Algorithmic Improvement and GPU Acceleration\\of the GenASM Algorithm
\vspace{-0.5em}
}

\author{\IEEEauthorblockN{Joël Lindegger}
\IEEEauthorblockA{\textit{ETH Zürich} \\
lijoel@ethz.ch}
\and
\IEEEauthorblockN{Damla Senol Cali}
\IEEEauthorblockA{
\textit{Bionano Genomics} \\
damlasenolcali@gmail.com}
\and
\IEEEauthorblockN{Mohammed Alser}
\IEEEauthorblockA{\textit{ETH Zürich} \\
alserm@ethz.ch}
\and
\IEEEauthorblockN{Juan Gómez-Luna}
\IEEEauthorblockA{\textit{ETH Zürich} \\
el1goluj@gmail.com}
\and
\IEEEauthorblockN{Onur Mutlu}
\IEEEauthorblockA{\textit{ETH Zürich} \\
omutlu@gmail.com}
}

\IEEEaftertitletext{\vspace{-3em}}
\maketitle

\begin{abstract}
We improve on GenASM, a recent algorithm for genomic sequence alignment, by significantly reducing its memory footprint and bandwidth requirement. Our algorithmic improvements reduce the memory footprint by 24\x\ and the number of memory accesses by 12\x. We efficiently parallelize the algorithm for GPUs, achieving a 4.1\x\ speedup over a CPU implementation of the same algorithm, a 62\x\ speedup over minimap2's CPU-based KSW2 and a 7.2\x\ speedup over the CPU-based Edlib for long reads.
\end{abstract}


\bstctlcite{IEEEexample:BSTcontrol}
\fontsize{10pt}{11pt}\selectfont
\vspace{-10pt}\section{Introduction}\vspace{-5pt}
Efficient algorithms for solving the genomic sequence alignment problem are based on \defn{dynamic programming (DP)}, such as the Smith-Waterman-Gotoh algorithm~\cite{GOTOH1982swgalgo}. These algorithms first construct a \defn{DP table} to find the best possible alignment score, followed by a \defn{traceback} step to retrieve the optimal alignment. They have quadratic time and space complexity~\cite{alser2020technology}, and asymptotically faster solutions are not to be expected~\cite{asm-lower-bound}. Hence, a significant effort has been and is being put towards speeding up this step through several approaches, such as prior filtering~(e.g. \cite{xin2015shiftedhammingdistance,alser2020sneakysnake,singh2021fpga,mansouri2022genstore}), constant factor algorithmic speedups~(e.g.~\cite{edlib, ksw2-a, ksw2-b}), GPU-based acceleration~(e.g.~\cite{gasal2,ahmed2020darwingpu}), FPGA-based acceleration~(e.g.~\cite{fei2018fpgasw}) or through specialized hardware accelerators~(e.g. \cite{cali2020genasm, turakhia2019darwin,alser2020accelerating}).

Our \textbf{goal}
is to speed up genomic sequence alignment over state-of-the-art software solutions for both short and long reads. To this end, we develop novel algorithmic improvements for the GenASM algorithm~\cite{cali2020genasm} and accelerate it using a GPU.

We choose the GenASM algorithm for its high throughput and fine-grained parallelism that makes it well suited for a GPU implementation. We observe that the GenASM algorithm exhibits high memory bandwidth pressure and its working set (DP table) does not fit into on-chip memory. We alleviate these limitations with three \textbf{key ideas}: We discover that (1)~the DP table can be \emph{compressed} by storing only the bitwise AND of the variables for each DP entry, (2)~part of the DP table can
opportunistically be \emph{excluded from calculation} if previous rows of the DP table already contain the full solution, and (3)~part of the DP table \emph{does not need to be stored} because the traceback operation cannot reach these entries. As a result, the \emph{entire} DP table fits into fast on-chip memory, and the number of accesses to the DP table is reduced significantly.

The \textbf{contributions} of this work are as follows:
\begin{itemize}
    \item We develop three novel algorithmic improvements to GenASM, collectively reducing the memory footprint by 24\x\ and the number of memory accesses by 12\x.
    \item Based on these insights, we develop CPU and GPU implementations of our improved GenASM algorithm that are capable of aligning both short and long reads.
    \item We demonstrate that our CPU and GPU implementations provide large speedups over the state-of-the-art sequence alignment software, KSW2~\cite{ksw2-a,ksw2-b} and Edlib~\cite{edlib}.
    \item We demonstrate that our algorithmic improvements are effective, and our GPU implementation efficiently parallelizes the improved GenASM algorithm.
\end{itemize}

\vspace{-5pt}\section{Results}\vspace{-5pt}
We evaluate the CPU implementation of our improved GenASM algorithm and the baseline sequence aligners (i.e. KSW2~\cite{ksw2-a,ksw2-b} and Edlib~\cite{edlib}) on a dual socket Intel Xeon Gold 5118 (2\x~24 logical cores) at 3.2GHz with 196GB DDR4 RAM, using 48 threads. We run our GPU implementation on an NVIDIA A6000~\cite{A6000}. We simulate 500 PacBio reads from the human genome using PBSIM2~\cite{pbsim2}, each of length 10 kb. We map these reads to the human genome using minimap2~\cite{ksw2-b} and obtain all chains (candidate locations) it generates using the \texttt{-P} flag, 138,929 locations in total. We align the (read,~reference) pairs obtained from the candidate locations using KSW2~\cite{ksw2-a,ksw2-b}, Edlib~\cite{edlib}, and both our CPU and GPU implementations.

Our CPU implementation achieves a 15.2\x, 1.7\x, and 1.9\x\ speedup over KSW2, Edlib, and a CPU implementation of GenASM without our improvements, respectively. Our GPU implementation achieves a 4.1\x, 62\x, 7.2\x, and 5.9\x\ speedup over our CPU implementation, KSW2, Edlib, and and a GPU implementation of GenASM without our improvements, respectively. We observe that the CPU and GPU implementations of GenASM provide speedups over Edlib \emph{only} if our algorithmic improvements are applied. We conclude that our algorithmic improvements are effective, and enable CPU and GPU implementations of GenASM to outperform state-of-the-art genomic sequence aligners.

\setstretch{0.9}
\def\refname{References\vspace{-5pt}}
\bibliographystyle{IEEEtran}
\vspace{-5pt}\bibliography{references}

\begin{thebibliography}{10}
\providecommand{\url}[1]{#1}
\csname url@samestyle\endcsname
\providecommand{\newblock}{\relax}
\providecommand{\bibinfo}[2]{#2}
\providecommand{\BIBentrySTDinterwordspacing}{\spaceskip=0pt\relax}
\providecommand{\BIBentryALTinterwordstretchfactor}{4}
\providecommand{\BIBentryALTinterwordspacing}{\spaceskip=\fontdimen2\font plus
\BIBentryALTinterwordstretchfactor\fontdimen3\font minus
  \fontdimen4\font\relax}
\providecommand{\BIBforeignlanguage}[2]{{%
\expandafter\ifx\csname l@#1\endcsname\relax
\typeout{** WARNING: IEEEtran.bst: No hyphenation pattern has been}%
\typeout{** loaded for the language `#1'. Using the pattern for}%
\typeout{** the default language instead.}%
\else
\language=\csname l@#1\endcsname
\fi
#2}}
\providecommand{\BIBdecl}{\relax}
\BIBdecl

\bibitem{GOTOH1982swgalgo}
O.~Gotoh, ``{An Improved Algorithm for Matching Biological Sequences},''
  \emph{Journal of Molecular Biology}, 1982.

\bibitem{alser2020technology}
M.~Alser \emph{et~al.}, ``{Technology dictates algorithms: Recent developments
  in read alignment},'' \emph{Genome Biology}, 2021.

\bibitem{asm-lower-bound}
A.~Backurs \emph{et~al.}, ``{Edit Distance Cannot Be Computed in Strongly
  Subquadratic Time (Unless SETH is False)},'' \emph{STOC}, 2015.

\bibitem{xin2015shiftedhammingdistance}
H.~Xin \emph{et~al.}, ``{Shifted Hamming distance: a fast and accurate
  SIMD-friendly filter to accelerate alignment verification in read mapping},''
  \emph{Bioinformatics}, 2015.

\bibitem{alser2020sneakysnake}
M.~Alser \emph{et~al.}, ``{SneakySnake: a fast and accurate universal genome
  pre-alignment filter for CPUs, GPUs and FPGAs},'' \emph{Bioinformatics},
  2020.

\bibitem{singh2021fpga}
G.~Singh \emph{et~al.}, ``{FPGA-Based Near-Memory Acceleration of Modern
  Data-Intensive Applications},'' \emph{IEEE Micro}, 2021.

\bibitem{mansouri2022genstore}
N.~Mansouri~Ghiasi \emph{et~al.}, ``{GenStore: a high-performance in-storage
  processing system for genome sequence analysis},'' \emph{ASPLOS}, 2022.

\bibitem{edlib}
M.~Šošić \emph{et~al.}, ``{Edlib: a C/C++ library for fast, exact sequence
  alignment using edit distance},'' \emph{Bioinformatics}, 2017.

\bibitem{ksw2-a}
H.~Suzuki \emph{et~al.}, ``Introducing difference recurrence relations for
  faster semi-global alignment of long sequences,'' \emph{BMC Bioinformatics},
  2018.

\bibitem{ksw2-b}
H.~Li, ``{Minimap2: pairwise alignment for nucleotide sequences},''
  \emph{Bioinformatics}, 2018.

\bibitem{gasal2}
N.~Ahmed \emph{et~al.}, ``{GASAL2: a GPU accelerated sequence alignment library
  for high-throughput NGS data},'' \emph{{BMC Bioinformatics}}, 2019.

\bibitem{ahmed2020darwingpu}
N.~Ahmed \emph{et~al.}, ``{GPU acceleration of Darwin read overlapper for de
  novo assembly of long DNA reads},'' \emph{BMC Bioinformatics}, 2020.

\bibitem{fei2018fpgasw}
X.~Fei \emph{et~al.}, ``{FPGASW: accelerating large-scale Smith--Waterman
  sequence alignment application with backtracking on FPGA linear systolic
  array},'' \emph{Interdiscip Sci}, 2018.

\bibitem{cali2020genasm}
D.~Senol~Cali \emph{et~al.}, ``{GenASM}: A high-performance, low-power
  approximate string matching acceleration framework for genome sequence
  analysis,'' \emph{MICRO}, 2020.

\bibitem{turakhia2019darwin}
Y.~Turakhia \emph{et~al.}, ``{Darwin: A Genomics Coprocessor},'' \emph{IEEE
  Micro}, 2019.

\bibitem{alser2020accelerating}
M.~Alser \emph{et~al.}, ``Accelerating genome analysis: a primer on an ongoing
  journey,'' \emph{IEEE Micro}, 2020.

\bibitem{A6000}
{NVIDIA}, ``{NVIDIA RTX A6000 Graphics Card},''
  https://www.nvidia.com/en-us/design-visualization/rtx-a6000, 2020.

\bibitem{pbsim2}
Y.~Ono \emph{et~al.}, ``{PBSIM2: a simulator for long-read sequencers with a
  novel generative model of quality scores},'' \emph{Bioinformatics}, 2020.

\end{thebibliography}
\end{document}